\documentclass[aps,prb,twocolumn,showpacs]{revtex4}
\usepackage{graphicx}

\begin{document}

\title{Fokker-Planck approach to the theory of magnon-driven spin Seebeck effect}
\author{L. Chotorlishvili$^1$, Z. Toklikishvili$^2$, V. K. Dugaev$^{1,3,4}$, J. Barna\'s$^{5,6}$, S. Trimper$^1$,
and J. Berakdar$^1$}

\affiliation{
$^1$Institut f\"ur Physik, Martin-Luther-Universit\"at Halle-Wittenberg,
Heinrich-Damerow-Str. 4, 06120 Halle, Germany\\
$^2$Department of Physics, Tbilisi State University, Chavchavadze av. 3, 0128, Tbilisi, Georgia\\
$^3$Department of Physics, Rzesz\'{o}w University of Technology,
al. Powstanc\'ow Warszawy 6, 35-959 Rzesz\'ow, Poland\\
$^4$Department of Physics and CFIF, Instituto Superior T\'ecnico,
TU Lisbon, Av. Rovisco Pais, 1049-001 Lisbon, Portugal\\
$^5$Faculty of Physics, Adam Mickiewicz University, ul. Umultowska 85, 61-614 Pozna\'n, Poland\\
$^6$Institute of Molecular Physics, Polish Academy of Sciences, ul. Smoluchowskiego 17, 60-179 Pozna\'n, Poland}

\begin{abstract}

Following the theoretical approach by Xiao {\it et al} [Phys. Rev. B 81, 214418 (2010)] to the spin Seebeck effect, we calculate
the mean value of the total spin current flowing through a normal-metal/ferromagnet interface. The
spin current emitted from the ferromagnet to the normal metal is evaluated in the framework of the Fokker-Planck
approach for the stochastic Landau-Lifshitz-Gilbert equation. We show that the total spin current depends not only
on the temperature difference between the electron and the magnon baths, but also on the external magnetic field and
magnetic anisotropy. Apart from this, the spin current is shown to saturate with increasing magnon temperature, and the saturation temperature
increases with increasing magnetic field and/or magnetic anisotropy.
\end{abstract}
\pacs{85.75.-d, 73.50.Lw, 72.25.Pn, 71.36.+c}
\date{\today }

\maketitle

\section{Introduction}

If two ends of a conductor are held at different temperatures, electrons from the hot
end diffuse towards the cold one.\cite{Ashcroft}
This phenomenon, discovered by Seebeck, is the basis for various
thermoelectric charge-transport effects and plays  a key role in the development of energy-saving technologies.
With the emergence of spintronics as a new area of mesoscopic physics, whose main objective is
to utilize the electron spin in device operations,  spin-related thermoelectricity has become of high interest.\cite{hatami}
Even though the generation of electromotive force by
temperature gradient has been known long ago, the spin analog of
the Seebeck effect, known as the spin Seebeck effect (SSE), was
discovered only very recently.\cite{Uchida} In the latter experiment, temperature gradient
along a ferromagnetic slab generated
a pure spin current over a long distance, much longer than in typical injection experiments, where spin
current (and also spin voltage) disappears over distances longer than the
spin-diffusion length.\cite{Uchida}
The SSE was observed not
only in metallic ferromagnets (like Co$_{2}$MnSi)\cite{Bosu} or
semiconducting ferromagnets (e.g. GaMnAs),\cite{Jaworski} but also
in magnetic insulators LaY$_{2}$Fe$_{5}$O$_{12}$ \cite{Xiao} and
(Mn,Zn)Fe$_{2}$O$_{4}$.\cite{Nonaka} Explanation
of the effect observed experimentally in insulating magnets cannot
rely on conduction electrons and requires a more general approach.

The Seebeck effect is usually  quantified by the Seebeck coefficient $S$
which is defined as the ratio of the generated electric voltage
$\Delta V$ to the temperature difference $\Delta T$, $\Delta V=-S\Delta T$.
The magnitude of the Seebeck coefficient $S$  depends on the scattering rate
and the density of electron states at the Fermi level, and thus it is
different in different materials. In the case of SSE, the spin voltage
is formally determined by
$ \mu_{\uparrow}- \mu_{\downarrow}$, where
$\mu_{\uparrow(\downarrow)}$ are the electrochemical
potentials for spin-up and spin-down electrons, respectively.
Usually, the density of states and the scattering rate for
spin-up and spin-down electrons are different, which results in different Seebeck constants for the two spin channels.
Therefore, in a metallic magnet subjected to a temperature
gradient, the electrons in different spin channels
generate different driving forces, leading to a spin voltage
that drives a nonzero
spin current.

In this paper we study the spin current flowing through
the normal-metal/ferromagnet interface due to the thermal bias applied to the system.
We consider the system and the model studied recently by Xiao {\it et al}.\cite{Bauer}
However, we use a different approach and also consider in detail the influence of an external magnetic field and of the magnetic anisotropy.
As in Ref.\onlinecite{Bauer}, we assume that the electron-phonon interactions in both normal-metal and ferromagnetic
subsystems are predominant,  as compared to the interface effects. Therefore, the phonon and  electron reservoirs in both
normal-metal and ferromagnet thermalize internally before the thermal equilibrium between
the ferromagnet and normal-metal appears.
In terms of the local temperature, which is  based on the hierarchy of relaxation times,
this means that the temperatures of the phonon ($T_{N(F)}^{p}$) and the electron ($T_{N(F)}^{e}$) baths are equal in both the normal-metal ($N$) and
the ferromagnet ($F$), $T_{N}^{p}=T_{N}^{e}=T_{N}$, $T_{F}^{p}=T_{F}^{e}=T_{F}$.
However, there is a difference in temperatures of the subsystems, $T_{N}\neq T_{F}$, which is externally controlled. This difference drives the SSE.
The interaction between the normal-metal and the ferromagnet subsystems is mediated {\it via} the magnon bath.
The temperature of the magnon bath deviates from the temperature of the electron and the phonon baths, $T_{F}^{m}\neq T_{N}$ and $T_{F}^{m}\neq T_{F}$.

As shown in Ref.\onlinecite{Bauer}, two different spin currents contribute to the total spin current flowing through the
normal-metal/ferromagnet interface. One of them is the spin current emitted from the ferromagnet to the normal metal due to the thermally activated magnetization dynamics in the ferromagnet.
This spin current is referred to as the spin-pump current, $\overrightarrow{I}_{sp}$.
The second contribution to the total spin current has the opposite nature and flows in the opposite direction -- from normal metal to the ferromagnet.
This contribution follows from the thermal noise in the normal metal and will be  referred to as the spin-torque current, $\overrightarrow{I}_{fl}$.
In order to evaluate the spin current flowing through the interface between normal metal and ferromagnet,
Xiao et al~\cite{Bauer} used the linearized  Landau-Lifshitz-Gilbert (LLG) equation and found that the spin current is
proportional to the difference in the temperatures of the magnon system
$T_{F}^{m}$ and electron subsystem in the normal metal $T^{e}_{N}$.
Here we address this problem using a different method, which is based on the Fokker-Planck equation for the stochastic
LLG  equation. We also distinguish between the influence of magnetic anisotropy and external magnetic field
on the spin Seebeck effect. As in the linearized approach the role of magnetic anisotropy is similar to that of an external magnetic field and can be described by some effective field, this is not the case when fluctuations are large.
We demonstrate that the spin current obtained within the framework of the linear response theory is
a particular case of the result obtained using the Fokker-Planck approach, and
corresponds to the low temperature approximation for the magnon temperature. Apart from this, we show that the spin current  saturates with increasing magnon temperature, and the saturation temperature
increases with increasing magnetic field and/or magnetic anisotropy.

The paper is organized as follows. In section II we describe the model. The Fokker-Planck equation is solved in section III, where two cases are distinguished: (i) the case with dominant external field, and (ii) the case where the uniaxial magnetic anisotropy field is dominant.
Summary and final conclusions are in section IV.

\section{General background}

We consider a ferromagnetic metallic layer which is in direct contact with a nonmagnetic metallic layer.
Magnetization dynamics of the ferromagnet will be
described by the LLG equation in the macrospin approximation.
\cite{Usadel, Evans, Kalmykov}
Following Xiao \emph{et al.},~\cite{Bauer} we assume that strong electron-phonon interaction assures
local thermal equilibrium between electrons and phonons
in both ferromagnetic and normal-metal layers,
$T_{F}^{p}=T_{F}^{e}=T_{F}$, $T_{N}^{p}=T_{N}^{e}=T_{N}$.
However, the
magnon temperature in the ferromagnetic layer is different from the corresponding temperature of
electrons, $T_{F}^{m}\neq T_{F}$. \cite{Bauer}

At finite temperatures, the thermally
activated magnetization dynamics in the ferromagnet gives rise to a
spin-current emitted from the ferromagnet to the normal
metal. This effect is known as the spin
pumping. \cite{Foros,Tserkovnyak}
The corresponding expression for the spin current density reads~\cite{Bauer}
\begin{equation}
\label{eq.1}
\vec{I}_{sp}=\frac{\hbar}{4\pi}\big[g_{r}\vec{m}(t)\times
\dot{\vec{m}}(t)+g_{i}\dot{\vec{m}}(t)\big],
\end{equation}
where $g_{r}$ and $g_{i}$ are the real and imaginary parts of the dimensionless spin
mixing conductance of the ferromagnet/normal-metal ($F|N$)  interface, while
$\vec{m}(t)$ is a dimensionless  unit vector along the magnetization direction.

In turn, the thermal noise in
the normal-metal layer leads to the spin current flowing from the normal metal to the ferromagnet,~\cite{Foros}
\begin{equation}
\label{eq.2}
\vec{I}_{fl}(t)=-\frac{M_{s}V}{\gamma}\gamma\vec{m}(t)\times\vec{h}'(t),
\end{equation}
where $M_{s}$ is the saturation magnetization,
$V$  is the total volume of the ferromagnet, and $\gamma$ is the gyro-magnetic factor. Apart from this,
$\vec{h}'(t)$ is the random magnetic field with the following
correlation function in the high temperature limit, $k_{B}T\gg
\hbar \omega_{0}$,
\begin{equation}\label{eq.3}
\langle\gamma h'_{i}(t)~\gamma h'_{j}(t')\rangle
=\sigma'^{2}\delta_{ij}\delta(t-t')
\end{equation}
for $i,j=x,y,z$.
Here, $\omega_{0}$ is the ferromagnetic resonance frequency,
$\sigma'^{2}=2\alpha'\gamma k_{B}T_{N}/M_{s}V$,
and $\alpha'=\gamma\hbar g_r/4\pi M_{s}V$ is
the magnetization damping constant related to the spin pumping.
Using Eqs~(\ref{eq.1}) to (\ref{eq.3}), the total average spin current flowing across the interface can be written in the form
\begin{equation}
\label{eq.4}
\langle\vec{I}_{s}\rangle=
\frac{M_{s}V}{\gamma}\big[\alpha'\langle\vec{m}\times\dot{\vec{m}}\rangle
+\gamma\langle\vec{m}\times\vec{h}'\rangle\big],
\end{equation}
while the magnetization dynamics is described by the stochastic
LLG equation,
\begin{equation}
\label{eq.5}
\dot{\vec{m}}=-\gamma\vec{m}\times\big(H_{\rm eff}\hat{z}+\vec{h}\big)+\alpha\vec{m}\times\dot{\vec{m}},
\end{equation}
where  $H_{\rm eff}$ is the
effective magnetic field which consists of the external constant magnetic field $H_{0}$ oriented along the $z$ axis
and  magnetic anisotropy field, $H_Am_z$, with $H_A=2K_1/M_s$ and $K_1$ being the anisotropy constant. For $K_1>0$ the magnetic anisotropy is of easy-axis type, while for $K_1<0$ it is of easy-plane type.  Apart from this, in the above equation $\hat{z}$ is the unit vector along the $z$ axis,  $\vec{h}$  is the total random field, while $\alpha$ is the total magnetic damping constant. \cite{Bauer}
This constant includes  the contributions from the bulk damping
constant $\alpha_{0}$  associated with the lattice random field
$h_{0}$  and from the damping constant $\alpha'$ associated with
the contact to the normal metal (random field $\vec{h}'(t)$).

We assume that the random contributions from the unrelated noise
sources are independent and therefore the correlation function for the
total random magnetic field can be  factorized in the following form:
\begin{equation}\label{eq.6}
\langle\gamma
h_{i}(t)~\gamma h_{j}(t')\rangle=\sigma^{2}\delta_{ij}\delta(t-t'),
\end{equation}
where $\sigma^{2}=2\alpha\gamma
k_{B}T_{F}^{m}/M_{s}V$, and
$\alpha T_{F}^{m}=\alpha_{0}T_{F}+\alpha'T_{N}$.

\section{Fokker-Planck equation and the spin current}

In Ref.~\onlinecite{Bauer}, the stochastic LLG equation was linearized near the relevant equilibrium.
Here,
to evaluate the mean current $\langle\vec{I}_{sp}\rangle$ for
the stochastic LLG, we derive the Fokker-Plank equation for the distribution
function $f(\vec{m},t)$.
The derivation procedure follows Ref.~\onlinecite{Risken} and is outlined in the Appendix.
As a result, one finds
\begin{eqnarray} \label{eq.7}
\frac{\partial f}{\partial t}&
=&\frac{1}{1+\alpha^2}\frac{\partial}{\partial \vec{m}} \bigg\{
(\vec{m}\times\vec{\omega}_{\rm eff})\, f +\alpha\vec{m}\times
\left( \vec{m}\times\vec{\omega}_{\rm eff}\right) f
\nonumber \\
&-&\frac{\sigma^2}{2(1+\alpha^2)}\; \vec{m}\times
\left( \vec{m}\times \frac{\partial f}{\partial\vec{m}}\right) \bigg\},
\end{eqnarray}
where $\vec{\omega}_{\rm eff}=\gamma H_{\rm eff}\hat{z}=(0,0,\omega_{\rm eff})$, with $\omega_{\rm eff}=\gamma H_{\rm eff}$.

The stationary solution of Eq. (\ref{eq.7}) for
the distribution function has the form
 \begin{eqnarray}\label{eq.8}
 f(\vec{m})&=&Z^{-1}\exp\left(\beta\int\vec{\omega}_{\rm eff}\cdot d\vec{m}\right),\\
 Z&=&\int\exp\left(\beta \int \vec{\omega}_{\rm eff}\cdot d\vec{m} \right)d^{3}
 \vec{m},\nonumber
 \end{eqnarray}
where we introduced the following notation: $\beta=2\alpha(1+\alpha^2)/\sigma^2 \approx 2\alpha/\sigma^2=M_{s}V/\gamma
 k_{B}T_{F}^{m}$. The limits of weak and strong magnetic anisotropy are of particular interest. Therefore, in the following we will consider both situations separately,
  and start with the case of a weak anisotropy field.

 \subsection{Weak magnetic anisotropy}

 In the case of weak magnetic anisotropy, when the external magnetic field is dominant,
 $\omega_{0}=\gamma H_{0}\gg \omega_{p}=\gamma H_{A}$, the effective field in Eq.~(\ref{eq.8}) is
 $\vec{\omega}_{\rm eff}=(0,0,\gamma H_{0})$. Using Eqs~(\ref{eq.1}) and (\ref{eq.8}) we find
 the mean values of the magnetization components
 \begin{eqnarray}\label{eq.9}
 && \langle m_{x}m_{y}\rangle=\langle m_{x}m_{z}\rangle=\langle
 m_{y}m_{z}\rangle=0,
 \nonumber \\
 && \langle m_{x,y}\rangle=0,
 \nonumber \\
 && \langle m_{z}\rangle=L(\beta \omega_{0}), \\
 &&  \langle m_{x,y}^{2}\rangle=\frac{1}{\beta \omega_{0}}L(\beta \omega_{0}),
 \nonumber \\
 && \langle m_{z}^{2}\rangle=1-\frac{2}{\beta \omega_{0}}L(\beta \omega_{0}),
 \nonumber
 \end{eqnarray}
where $L(x)=\coth x-\frac{1}{x}$ is the Langevin function, which has the
asymptotics $L(x)\approx x/3$, $x\ll 1$.
Then, taking into account Eq.~(\ref{eq.9}), one finds the mean value of the
spin current,
\begin{equation}\label{eq.10}
\langle I_{sz}\rangle=\frac{M_{s}V}{\gamma}\big\{\alpha' \omega_{0}(1-\langle
m_{z}^{2}\rangle)-\gamma \langle (m_{x}h'_{y}-m_{y}h'_{x})\rangle
\big\}
\end{equation}
The last term in Eq.~(\ref{eq.10}) can be evaluated using the
linear-response theory, i.e., by linearizing the LLG equation  in the vicinity of
the equilibrium point, $\langle m_{z}\rangle=L(\beta \omega_{0})$.
After straightforward but laborious calculations one obtains
\begin{equation}\label{eq.11}
\langle
m_{x}h'_{y}-m_{y}h'_{x}\rangle=\frac{\sigma'^{2}}{\gamma}\langle
m_{z}\rangle.
\end{equation}
Combining Eqs. (\ref{eq.10}) and (\ref{eq.11}) one can write
the final result for the spin current density in the form
\begin{equation}\label{eq.12}
\langle I_{sz}\rangle=2 \alpha'
k_{B}L\bigg(\frac{M_{s}VH_{0}}{k_{B}T_{F}^{m}}\bigg)
\big(T_{F}^{m}-T_{N}\big).
\end{equation}

We see that the average spin current depends on two physical
parameters: (i) the ratio of the  magnetic energy in the external field to the
thermal energy corresponding to the magnon temperature, $M_sVH_{0}/k_BT_{F}^{m}$, and (ii) the difference between
the magnon temperature and the temperature of the electron-phonon bath in
the normal metal, $\big(T^{m}_{F}-T_{N}\big)$.
The dependence of the average spin
current on the field is factorized in the Langevin
function, Eq.(\ref{eq.12}). Therefore, introducing the notation $\langle
I_{sz}\rangle_{0}$ for the mean spin current calculated in the
linear response approach, $\langle I_{sz}\rangle_{0}=2 \alpha'
k_{B} \big(T^{m}_{F}-T_{N}\big)$,\cite{Bauer} one can rewrite Eq.(\ref{eq.12})
in the compact form as  $\langle
I_{sz}\rangle=L\bigg(\frac{M_{s}VH_{0}}{k_{B}T_{F}^{m}}\bigg)\langle
I_{sz}\rangle_{0}$.

In the limit of a low magnon temperature,
$H_{0}/T_{F}^{m}>\frac{k_{B}}{M_{s}V}$, we have $\langle I_{sz}\rangle\approx\langle
I_{sz}\rangle_{0}$. This means that the spin current calculated using
the Fokker-Plank approach, without a linearization of the system, gives
the same result as that obtained in the linear
response. The physical reason for this is
clear. Indeed, in the case of a strong magnetic field, the magnetization vector
tends to be aligned along the field direction, and nonlinear
effects in the magnetization dynamics related to large deviation
from the equilibrium are less relevant. However, in the
opposite case, corresponding to the high magnon temperature,
$H_{0}/T_{F}^{m}<\frac{k_{B}}{M_{s}V}$, and strong magnetization fluctuations,
the nonlinear effects in the magnetization dynamics are much more
important.  Consequently, the spin current is different
from $\langle I_{sz}\rangle_{0}$ and reads $\langle I_{sz}\rangle
=\frac{2}{3}\alpha'M_{s}H_{0}\big(1-T_{N}/T_{F}^{m}\big)$. We see
that the maximum value of the spin current corresponds to the hot magnon
bath and saturates at  $\langle I_{sz}\rangle
=\frac{2}{3}\alpha' M_{s}H_{0}$. In turn, the spin current
from linear response theory increases linearly with the
magnon temperature.

\begin{figure}[h]\label{fig.1}
\includegraphics*[width=0.5\textwidth]{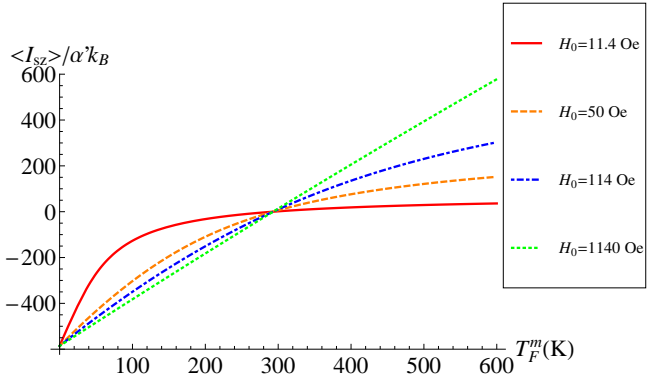}
\caption{Dependence of the spin current $\big<I_{sz}\big>$ on
magnon temperature $T_{F}^{m}$ for the following
parameters:\cite{Bauer} $M_{s}=800$~G, $V=1.6\times 10^{-18}$~cm$^{3}$,
$T_{N}=293$~K. The parameters correspond to
Py=Ni$_{80}$Fe$_{20}$ alloy.\cite{Bauer} The spin current is measured
in units of ${\alpha}^\prime k_{B}$ and is shown for different amplitudes of the magnetic field.
The dotted (green) line corresponds to the linear response theory.}
\end{figure}

\begin{figure}[h]\label{fig.2}
\includegraphics*[width=0.5\textwidth]{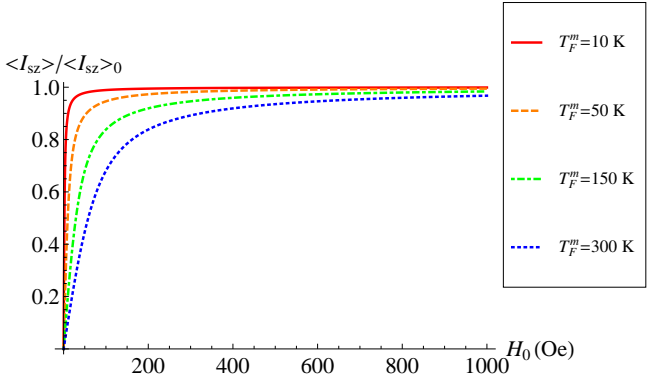}
\caption{Dependence of the ratio
$\big<I_{sz}\big>/\big<I_{sz}\big>_{0}$ on the magnetic field
$H_{0}$  for the following  parameters:\cite{Bauer}
$M_{s}=800 G$, $V=1.6\times 10^{-18} cm^{3}$. The parameters
correspond to the Py=Ni$_{80}$Fe$_{20}$ alloy.\cite{Bauer} Different curves correspond to different
values of the magnon temperature.}
\end{figure}

The dependence of the spin current on the magnetic field and
magnon temperature is plotted in Figs.~1 and 2. In particular, in Fig.~1 the
dependence of the total spin current on the magnon temperature is plotted for different values of the
external magnetic field. The dotted (green) line corresponds to the linear response theory, whereas all other curves
correspond to the Fokker Planck approach. We note that in contrast to the linear response theory,
the Fokker-Planck approach leads to the spin current
that saturates  for high magnon temperatures. This, in turn,  means
that the linear response is valid only in a narrow range of the magnon temperatures.
Note, that all the curves  cross  the same point for $T^m_F=0$ as there are then no thermal fluctuations in the magnon system, and thus
the only contribution comes from thermal noise in the nonmagnetic film. Deviation from the description based on the linearized model is especially large
for small magnetic fields and appears already at low magnon temperatures (see the curve for $H_0=11.4$ Oe in Fig.1). In this case, the saturation of the
spin current also appears at low temperatures. For higher magnetic field, the deviation is smaller and appears at higher magnon temperatures. This behavior is
reasonable as the fluctuations at low magnetic fields are larger, and therefore difference between linearized model and description based on the Fokker Planck equation appears at lower magnon temperatures. Obviously, the curves also cross at the point corresponding to $T^m_F=T_N$, where the spin current vanishes. This directly follows from Eq.(\ref{eq.12}). In  the nonlinear regime
$L\bigg(\frac{M_{s}VH_{0}}{k_{B}T_{F}^{m}}<1\bigg)\approx\frac{M_{s}VH_{0}}{3k_{B}T_{F}^{m}}$ and for spin current we have
$\langle I_{sz}\rangle=
\frac{2\alpha'M_{s}VH_{0}}{3T_{F}^{m}}\big(T_{F}^{m}-T_{N}\big)$. Consequently  the magnon temperature $T_{F}^{m}$ for which we observe saturation
of the spin current $\langle I_{sz}\rangle=\frac{2}{3}\alpha'M_{s}VH_{0}$ basically is defined by the following inequalities
$T_{F}^{m}>\frac{M_{s}VH_{0}}{k_{B}},~~T_{F}^{m}>T_{N}$.

Figure 2 shows the dependence of the ratio $\big<I_{sz}\big>/\big<I_{0}\big>$ on the
magnetic field for different values of the magnon temperatures $T_{F}^{m}$. This figure clearly shows that the saturation field increases with
increasing the magnon temperature. Apart from this, it is interesting to note, that $\big<I_{sz}\big>/\big<I_{0}\big>$  tends to zero in the limit of zero magnetic field. This behavior can be accounted for by noting that $\omega_0\to 0$ for zero magnetic field, so the zero-field magnetic fluctuations are large. In this limit the description based on the linearized model gives finite spin current, while that based on the Fokker Planck equation gives vanishing spin current.

\subsection{Strong magnetic anisotropy}

In the presence of a
magnetic anisotropy, the situation is more complicated.
Now,   $\vec{\omega}_{\rm eff}=\omega_{\rm eff}(m_{z})\hat{z}$, where
$\omega_{\rm eff}(m_{z})=\gamma H_{\rm eff}=\gamma(H_{0}+H_{A}m_{z})$.  The stationary solution to  equation (\ref{eq.7}) is
\begin{equation}\label{eq.14}
f=Z_{a}^{-1}\exp\bigg[\frac{2\alpha}{\sigma^2}\bigg(\omega_{0}m_{z}
+\frac{\omega_{p}m_{z}^2}{2}\bigg)\bigg],
\end{equation}
where
$Z_{a}=\int\exp\bigg[\frac{2\alpha}{\sigma^2}\bigg(\omega_{0}m_{z}
+\frac{\omega_{p}m_{z}^2}{2}\bigg)\bigg]d^{3}\vec{m}$
is a normalization factor, and $\omega_{p}=\gamma H_{A}$. Using Eq.(\ref{eq.14})
and calculating the spin current we find
\begin{eqnarray}\label{eq.15}
&& \langle I_{sz}\rangle=\alpha' k_{B}\bigg\{T_{F}^{m}A(1-\langle
m_{z}^{2}\rangle)+\\
&& +2T_{F}^{m}B(\langle m_{z}\rangle-\langle
m_{z}^{3}\rangle)-2T_{N}\langle m_{z}\rangle \bigg\}.\nonumber
\end{eqnarray}
Here, the mean values $\langle
m_{z}\rangle$, $\langle m_{z}^{2}\rangle$, and $\langle
m_{z}^{3}\rangle$ are given by
\begin{eqnarray}\label{eq.16}
&&\langle
m_{z}\rangle=\frac{e^{A}\sinh(A)}{\sqrt{B}G(A,B)}-\frac{A}{2B};\nonumber \\
&&\langle
m_{z}^{2}\rangle=-\frac{1}{2B}+\bigg(\frac{A}{2B}\bigg)^{2}+\\
&&+\frac{e^{A}(2B\cosh(A)-A\sinh(A))}{2B^{3/2}G(A,B)};\nonumber\\
&&\langle
m_{z}^{3}\rangle=\frac{3}{4}\frac{A}{B^2}-\bigg(\frac{A}{2B}\bigg)^{3}+\nonumber\\
&&+\frac{e^{A}\big(-2AB\cosh
A+(A^2-4B-4B^2)\sinh(A)\big)}{4B^{5/2}G(A,B)}.\nonumber
\end{eqnarray}
The following notation has been  introduced in the above equations:
$A=\frac{M_{s}VH_{0}}{k_{B}T_{F}^{m}}$,
$B=\frac{K_{1}V}{k_{B}T_{m}^{F}}$, and
$G(A,B)=e^{2A}F\bigg(\frac{A+2B}{\sqrt{B}}\bigg)-F\bigg(\frac{A-2B}{\sqrt{B}}\bigg)$, where
$F(x)=\exp(-x^2)\int\limits_{0}^{x}\exp(y^2)dy$ is the Dawson
function.\cite{Abramovitz}
The above formula,  Eqs.~(\ref{eq.15}) and
(\ref{eq.16}), will be now used to calculate the influence of magnetic anisotropy on the
asymptotic behavior of the average spin current.

The effect of magnetic anisotropy
is demonstrated in Fig.~3, where the spin current is shown as a function of the external field for the indicated values of $H_A$.
It is evident, that for the parameters assumed in Fig.3 the  magnetic anisotropy
magnifies the spin current for $H_A>0$ (easy axis anisotropy).  This behavior is qualitatively similar to that observed for external field $H_0$. The difference comes from the fact that the effective role of anisotropy depends on the magnetization -- the anisotropy field changes sign when the $m_z$ component is reversed.   One may also say, that the effect of magnetic anisotropy adds to the effect of magnetic field. In turn, the easy plane anisotropy ($H_A<0$) reduces the spin current, i.e. reduces the effect of external field.  Generally, the spin current saturates with increasing $H_0$. However, in the presence of easy-axis (easy-plane) magnetic anisotropy, the
saturation is reached at $H_0$ lower (larger) than in the absence of the magnetic anisotropy.

\begin{figure}[h]\label{fig.3}
\includegraphics*[width=0.5\textwidth]{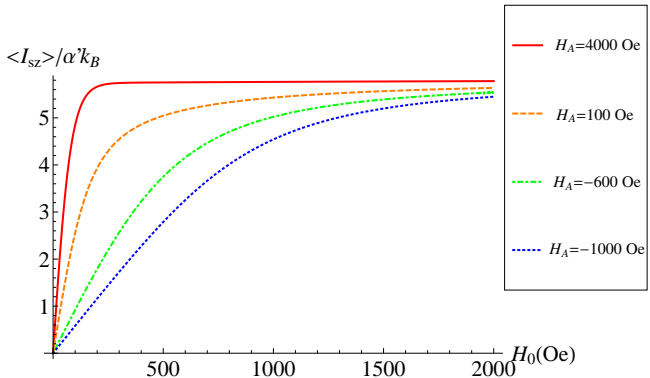}
\caption{The dependence of spin current $\langle I_{sz}\rangle$ on the
field $H_{0}$ for
the following parameters \cite{Bauer}: $T^{m}_{F}=293
K $, $T_{ N}=290.07 K $, $\Delta
T=T^{m}_{F}-T_{N}=2.93 K$, $4\pi M_{s}=4000$~G, $V=1.6\times
10^{-18}$~cm$^{3}$. These parameters correspond to manganese
spinel ferrite films \cite{Xu} (MnFe$_{2}$O$_{4}$). The spin current is
measured in the units of ${\alpha}^\prime k_{B}$. Different curves correspond to different values of $H_A$.}
\end{figure}

\section{Summary and conclusions}

We have studied the spin Seebeck effect in a
system consisting of normal-metal and ferromagnetic films subjected to a temperature
gradient. Using the Fokker-Planck equation we  have derived analytical expressions for the averaged spin current flowing
through the interface between the  layers. This current consists of two parts:
The first part is the spin current which occurs due to the thermally activated magnetization dynamics
in the ferromagnetic layer, $\big<I_{sp}\big>$.
The second part, $\big<I_{fl}\big>$, flows in the opposite direction and
arises from thermal fluctuations in the normal metal.

We have considered two special cases -- with and without magnetic anisotropy.
The obtained analytical results describe the dependence of the mean spin current on the external magnetic field, magnetic
anisotropy field,  and on the
difference between the magnon temperature and the temperature of the electron-phonon bath in the normal metal.
The dependence  of the spin current on the thermal gradient was already analyzed in Ref.\onlinecite{Bauer},
where the corresponding stochastic Landau-Lifshitz-Gilbert equation was linearized.
However, the dependence on magnetic anisotropy  was not considered there.
We have shown that the magnetic field enhances the spin current, which  should be observable  experimentally.
In the absence of the anisotropy, the dependence of spin current on the magnetic field is factorized
by the Langevin function, Eq.~(\ref{eq.12}).
In the limit of a low magnon temperature, $H_{0}/T_{F}^{m}>\frac{k_{B}}{M_{s}V}$,
the spin current calculated using the approach based on the Fokker-Plank equation (without linearization
of the Landau-Lifshitz-Gilbert equation) gives the same result as that obtained in the framework of linear response theory.
From the physical point of view this result is rather clear.
In case of a strong magnetic field, the magnetization vector tends to be aligned along the field
and the nonlinear effects in the magnetization dynamics concerning large deflection from
the equilibrium position are less relevant. However, in the opposite case corresponding to the
high magnon temperature $H_{0}/T_{F}^{m}<\frac{k_{B}}{M_{s}V}$ and larger fluctuations of the magnetization vector,
nonlinear effects in the magnetization dynamics are more important.
Consequently, the behavior of the spin current is different from that found in the linear response description.

\section{Acknowledgements}

The financial support by the Deutsche Forschungsgemeinschaft (DFG)
through SFB 762, contract BE 2161/5-1, is gratefully acknowledged.
This work was supported by the National Science
Center in Poland as the Project No. DEC-2012/04/A/ST3/00372 and as
a research project in years 2011 -- 2014. This project is also
financially supported by the Shota Rustaveli National Science
Foundation under Grant No.~30/12.

 \appendix
 \section{Derivation of the Fokker-Plank equation} \label{App:AppendixA}
 For the derivation of the Fokker-Plank equation we  follow
Ref.\cite{Risken} and use the functional integration method in
order to average the dynamics over all possible realizations of the
random noise field. First, we rewrite eq.(\ref{eq.5}) in the
following form:
\begin {eqnarray} \label{eq.A1}
\dot{\vec{m}}=&-&\frac{1}{1+\alpha^2}\vec{m}\times(\vec{\omega}_{eff}+\vec{\zeta}(t))-\\
&-&\frac{\alpha}{1+\alpha^2}\vec{m}\times(\vec{m}\times\vec{\omega}_{eff}),
\nonumber
\end{eqnarray}
where $\vec{\omega}_{eff}=\gamma
H_{eff}\hat{z}=(0,0,\omega_{eff})$, and
$\vec{\zeta}(t)=\gamma\vec{h}(t)$ is a random Langevin field with
the following correlation relations:
\begin{eqnarray}\label{eq.A2}
&&\langle \vec{\zeta}(t)\rangle=0  \\
&& \langle \zeta_{i}(t); \zeta_{j}(t')\rangle=\frac{2\alpha \gamma
k_{B}T_{m}^{F}}{M_{s} V} \delta_{ij} \delta(t-t')\equiv \sigma^2
\delta_{ij} \delta(t-t').\nonumber
\end{eqnarray}
We introduce the probability distribution for the random Gaussian
noise $\vec{\zeta}$:
\begin{equation}\label{eq:A3}
F[\vec{\zeta}(t)]=\frac{1}{Z_{\zeta}}\exp\bigg[-\frac{1}{\sigma^2}\int\limits_{-\infty}^{+\infty}d\tau
\zeta^{2}(\tau)\bigg],
\end{equation}
where $Z_{\zeta}=\int D\vec{\zeta} F$ is the noise partition
function,  $D\vec{\zeta}$ denotes the functional integration over all
realizations of $\vec{\zeta}(\tau)$. For all $n$ we have:
\begin{equation} \label{eq:A4}
\int
D\vec{\zeta}\frac{\delta^{n}F[\vec{\zeta}]}{\delta\zeta_{\alpha_{1}}\delta\zeta_{\alpha_{2}}\ldots
\delta\zeta_{\alpha_{n}}}=0.
\end{equation}
Using  Eq.(\ref{eq:A3}), the average of any noise the functional
$A[\vec{\zeta}(t)]$ can be written as
\begin{equation}\label{eq:A5}
\langle A[\vec{\zeta}]\rangle_{\zeta}=\int D \vec{\zeta}
A[\vec{\zeta}]F[\vec{\zeta}].
\end{equation}
Using the identity $\frac{\delta \zeta_{\alpha}(\tau)}{\delta
\zeta_{\beta}(t)}=\delta_{\alpha \beta}\delta(\tau-t)$ and
(\ref{eq:A4})-(\ref{eq:A5}) for $n=1,2$, it is easy  to obtain the
correlation relations (\ref{eq.A2}).\cite{Risken} The Fokker-Plank
equation corresponding to Eq.(\ref{eq.A1}) can be written for the
distribution function as
\begin{equation}\label{eq.A6}
f(\vec{M},t)\equiv\langle \vec{\pi} (t,
[\vec{\zeta}])\rangle_{\zeta},~~~ \vec{\pi}(t,[\vec{\zeta}])\equiv
\delta (\vec{M}-\vec{m}(t))
\end{equation}
on the sphere $|\vec{M}|=1$. Taking into account the
relation\cite{Risken} $\dot{\vec{\pi}}=-\frac{\partial \vec{\pi}
}{\partial \vec{M}}\dot{\vec{m}}$ and the equation of motion
(\ref{eq.A1}) we deduce  the following Fokker-Plank equation
\begin{eqnarray} \label{eq.A7}
\frac{\partial f}{\partial
t}&=&\frac{1}{1+\alpha^2}\frac{\partial}{\partial
\vec{m}}\bigg\{(\vec{m}\times\vec{\omega}_{eff})f+[\vec{m}\times
(\vec{m}\times\vec{\omega}_{eff})]f+
\nonumber \\
 &+& \vec{m}\times\langle \vec{\zeta}(t)\vec{\pi}(t,[\vec{\zeta}])\rangle \bigg\}.
\end{eqnarray}
To calculate  $\langle
\vec{\zeta}(t)\vec{\pi}(t,[\vec{\zeta}])\rangle$ we use the standard
procedure\cite{Risken}, which yields
\begin{equation}\label{eq.A8}
\langle
\vec{\zeta}(t)\vec{\pi}(t,[\vec{\zeta}])\rangle=-\frac{\sigma^2}{2(1+\alpha^2)}\vec{m}\times\frac{\partial
f}{\partial \vec{m}}.
\end{equation}
Inserting Eq.(\ref{eq.A8}) into Eq.(\ref{eq.A7}) we find the Fokker-Planck
equation, Eq.(\ref{eq.7}).


\end{document}